\documentclass[12pt]{article}
\usepackage{epsf}
\def\be{\begin{equation}}
\def\ee{\end{equation}}
\def\bea{\begin{eqnarray}}
\def\eea{\end{eqnarray}}

\begin{document}
\begin{titlepage}
\begin{center}
{\Large \bf William I. Fine Theoretical Physics Institute \\
University of Minnesota \\}
\end{center}
\vspace{0.2in}
\begin{flushright}
FTPI-MINN-14/39 \\
UMN-TH-3410/14 \\
November 2014 \\
\end{flushright}
\vspace{0.3in}
\begin{center}
{\Large  $\chi_{c0} \, \omega$ production in $e^+e^-$ annihilation through $\psi(4160)$.
\\}
\vspace{0.2in}
{\bf Xin Li$^a$  and M.B. Voloshin$^{a,b,c}$  \\ }
$^a$School of Physics and Astronomy, University of Minnesota, Minneapolis, MN 55455, USA \\
$^b$William I. Fine Theoretical Physics Institute, University of
Minnesota,\\ Minneapolis, MN 55455, USA \\
$^c$Institute of Theoretical and Experimental Physics, Moscow, 117218, Russia
\\[0.2in]

\end{center}

\vspace{0.2in}

\begin{abstract}
We argue that the recent BESIII data on the cross section for the process $e^+e^- \to \chi_{c0} \, \omega$  in the center of mass energy range 4.21 - 4.42 GeV can be described by the contribution of the known charmonium-like resonance $\psi(4160)$ with the mass of about 4190\,MeV. The value of the coupling in the transition $\psi(4160) \to \chi_{c0} \, \omega$ needed for this mechanism is comparable to that in another known similar transition  $\chi_{c0}(2P) \to J/\psi \, \omega$. The suggested mechanism also naturally explains the reported relative small value of the cross section for the final states $\chi_{c1} \, \omega$ and $\chi_{c2} \, \omega$ above their respective thresholds.
\end{abstract}
\end{titlepage}

The recently reported~\cite{bes} BESIII data on production of the final states $\chi_{cJ} \, \omega$ in the $e^+e^-$ annihilation at $\sqrt{s}$ from 4.21 to 4.42 GeV indicate a peak in the cross section for  $\chi_{c0} \, \omega$ at about 4.23 GeV and apparently no corresponding peaks above the thresholds for $\chi_{cJ} \, \omega$ with $J=1$ and $J=2$. The best fit to the $\chi_{c0} \, \omega$ data with a resonance curve yields the parameters of the resonance\cite{bes}: $M=(4230 \pm 8)$\,MeV and $\Gamma = (38 \pm 12)$\,MeV, which parameters do not correspond to any of the previously known charmonium-like states. It is not unusual recently that new quarkonium-like resonances are revealed in various channels with a hidden heavy flavor, and the newly observed peak could indicate an existence of another such state. Here, however, we explore a somewhat more routine interpretation of the observed peak as being due to a well known charmonium $J^{PC}=1^{--}$ resonance, namely the $\psi(4160)$. Despite the notation $\psi(4160)$ the actual mass of this state is in fact~\cite{pdg} $M_\psi = (4191 \pm 5)$\,MeV (and $\Gamma_\psi = (70 \pm 10)$\,MeV. The shift in the mass from the initial data to the current higher value is mostly due to the re-analysis~\cite{bes2} of the interference with the $\psi(4040)$ peak.  We argue that a close proximity of the actual mass of $\psi(4160)$ to the observed enhancement of the $e^+e^- \to \chi_{c0} \, \omega$ cross section makes our interpretation well compatible with the data. The suppression of the production of $\chi_{c1}$ and $\chi_{c2}$ in similar processes above their thresholds as well as of $\chi_{c0}$ in that range of higher energy is then merely due to larger distance from the resonance peak.

If confirmed by the future data, the discussed interpretation would imply that the process $e^+e^- \to \chi_{c0} \, \omega$ is not as much of direct relevance to the searches for new charmonium-like states, possibly of a complex structure, but rather falls into another very interesting class of hadronic transitions between quarkonium states, more specifically, of the transitions with the emission of the $\omega$ meson. Such transitions between the $J^{PC}=1^{--}$ and the $\chi_J$ states were observed in both charmonium: $\chi_{c0}(2P) \to J/\psi \, \omega$~\cite{belle}, and bottomonium: $\chi_{b1,b2}(2P) \to \Upsilon(1S) \, \omega$~\cite{cleo}, and most recently~\cite{belle2} at the $\Upsilon(5S)$ resonance: $\Upsilon(5S) \to \chi_{b1} \, \omega$ (and also an indication of $\Upsilon(5S) \to \chi_{b2} \, \omega$). These processes are not suppressed by any (approximate) symmetries in QCD and are allowed to proceed in the $S$ wave~\footnote{Within the multipole expansion in QCD~\cite{gottfried,mv0} these processes arise in the third order in the leading $E1$ interaction~\cite{mv}. This illustrates  absence of a suppression, even though the multipole expansion is hardly applicable at a quantitative level to the processes with highly excited quarkonium states.} . In the limit of exact heavy quark spin symmetry the transitions for the $\chi_{J}$ states with different $J$ are related by the generic expression~\cite{mv} for the $S$-wave amplitude:
\be
A(\psi \chi_{J} \omega) = g_\omega \, \left ( \psi_i \omega_i \chi_0 + \sqrt{3 \over 2} \, \epsilon_{ijk} \psi_i \omega_j \chi_k + \sqrt{3} \, \psi_i \omega_j \chi_{ij} \right )~,
\label{amp}
\ee
where the nonrelativistic limit for heavy quarkonium is assumed with the nonrelativistic normalization for the heavy states, $\vec \psi$ and $\vec \omega$ stand for the polarization amplitudes for the $1^{--}$ state ($\psi$) and the $\omega$ meson, and $\chi_0$, $\vec \chi$ and $\chi_{ij}$ are the amplitudes for the $\chi_J$ states. The dimensionless constant depends on the specific $1^{--}$ state and on the considered multiplet of the $\chi_J$ states. In our normalization the rate of e.g. the decay $\psi(4160) \to \chi_{c0} \, \omega$ is given by
\be
\Gamma[\psi(4160) \to \chi_{c0} \, \omega] = g_\omega^2 {p_\omega \over 2 \pi}
\label{gam}
\ee
with $p_\omega$ being the momentum of the emitted $\omega$.

If one neglects the widths of the $\omega$ and $\chi_{c0}$ resonances and approximates the $\psi(4160)$ resonance by a simple Breit-Wigner shape, the cross section for the process $e^+e^- \to \psi(4160) \to \chi_{c0} \, \omega$ is given by
\be
\sigma (e^+e^-  \to \chi_{c0} \, \omega) = g_\omega^2 \, {3 \over 2 M^2} \, {\Gamma_{ee} \, p_\omega(E) \over (E-M)^2 + \Gamma^2/4}~
\label{sig}
\ee
where $E=\sqrt{s}$ is the center of mass energy, $M$ and $\Gamma$ are the mass and the total width of the $\psi(4160)$ resonance, and $\Gamma_{ee}$ is its partial width of decay to $e^+e^-$ currently measured with a considerable uncertainty~\cite{pdg}, $\Gamma_{ee} = (0.48 \pm 0.22)$\,keV. Asuuming the central value for the latter width, the expression in Eq.(\ref{sig}) reproduces the experimentally measured cross section of approximately 55\,pb at $E=4230$\,MeV, at which point the largest statistics is available and the experimental errors are the smallest, if $g_\omega^2 \approx 4 \times 10^{-2}$. Naturally, the uncertainty in this estimate is large, about 50\%, mainly due to poor knowledge of the $\Gamma_{ee}$.

For a more detailed fit to the data of the shape of the energy dependence of the cross section described by the discussed resonance mechanism, especially at the lower end of the relevant energy range, we have included the effects of  the finite widths of the $\omega$ resonance (8.5\,MeV) and of the $\chi_{c0}$ charmonium state (10.5\,MeV). Since we neglect any possible non-resonant background and any variation of the width of $\psi(4160)$ at energies well above the resonance, the only parameter in the fit is the overall normalization, i.e. the coupling $g_\omega^2$. A comparison of our fit with the data is shown in Fig.~1. The quality of the fit is $\chi^2/N_{d.o.f.} = 9.9/8$, which although is likely worse than the fit to a new resonance in Ref.~\cite{bes}, but is still compatible with the data within one standard deviation. It can be noted that most likely the simple Breit-Wigner approximation has to be modified at the higher end of the measured energy range. Thus if we fit only the data at 4.31\,GeV and below (i.e. not including the three highest energy points) the  figure of merit for our fit improves to $\chi^2/N_{d.o.f.} = 4.0/5$.

\begin{figure}[ht]
\begin{center}
 \leavevmode
    \epsfxsize=13cm
    \epsfbox{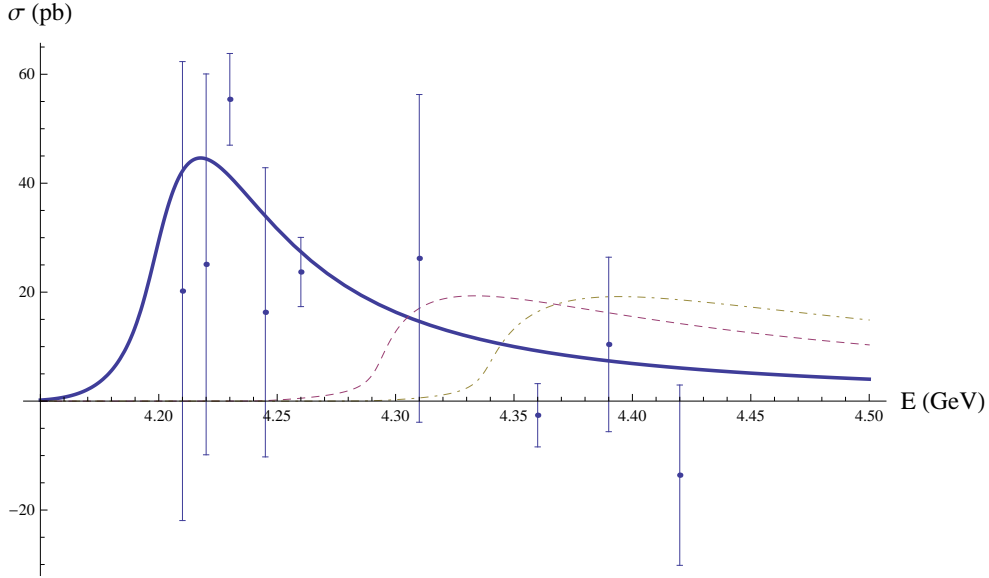}
    \caption{The fit of the $\psi(4160)$ resonance contribution to the data~\cite{bes} for $\sigma(e^+e^- \to \chi_{c0} \, \omega)$ (solid). Also shown are the curves for the cross section in the channels $\chi_{c1} \, \omega$ (dashed) and $\chi_{c2} \, \omega$ (dot-dash). }
\end{center}
\end{figure} 

Also shown in Fig.~1 are the expected from the formulas in Eqs.~(\ref{amp}) and (\ref{sig}) curves for the cross section in the channels $\chi_{c1} \, \omega$ and $\chi_{c2} \, \omega$. It should be noted however that a straightforward application of these formulas may suffer from relatively large modifications. Such modifications can arise both from a deviation at higher energies of the resonance curve from the Breit-Wigner approximation, and from possible other non-resonant contributions to the production mechanism, e.g. mediated by the heavy meson-antimeson pairs~\cite{mv2,mllz}. In particular, the latter contributions can potentially violate the heavy quark symmetry prediction for the ratio of the yield $\sigma/p_\omega$ in the channels $\chi_{cJ} \, \omega$ with different $J$: $\chi_{c0}:\chi_{c1}:\chi_{c2} = 1:3:5$.

We have also examined a possible contribution of the higher charmonium-like resonance $\psi(4415)$. The data~\cite{bes} indicate no enhancement at the energy of this state. Neither our fit showed any improvement with inclusion of this additional resonance. We find that its contribution to the production of $\chi_{c0} \, \omega$ should be less than 10\% of that of $\psi(4160)$ in the amplitude. This probably indicates that the resonances $\psi(4160)$ and $\psi(4415)$ have significantly different internal dynamics.

In order to assess how reasonable is the value of the coupling $g_\omega^2 \sim 10^{-2}$ needed for the discussed interpretation to work, one may turn to the only known another similar process in charmonium, the decay $\chi_{c0}(2P) \to J/\psi \, \omega$. Using the notation $g_\omega(2P \to 1S)$ for the coupling, normalized as in Eq.(\ref{amp}), the rate of the decay can be written as
\be
\Gamma[\chi_{c0}(2P) \to J/\psi \, \omega] = g^2_\omega(2P \to 1S) \, {3 \, p_\omega \over 2 \pi} \approx \left ( {g^2_\omega(2P \to 1S) \over 10^{-2}} \right ) \, 1\,{\rm MeV}~.
\label{gam2}
\ee
This width is currently unknown from direct measurements, but it can be argued that it is likely to be larger than 1\,MeV. Indeed, the total width is measured as $\Gamma[\chi_{c0}(2P)]=(20 \pm 5)$\,MeV, and also~\cite{pdg} $\Gamma[\chi_{c0} \to \gamma \gamma] \times Br[\chi_{c0}(2P) \to J/\psi \, \omega] = (54 \pm 9)$\,eV. The two photon decay rate of a $P$-wave state in the potential model is proportional to the square of the derivative of the radial function at the origin $|R'_P(0)|^2$ and should be significantly smaller for the the $2P$ state $\chi_{c0}(2P)$ than  for the $1P$, $\Gamma(\chi_{c0} \to \gamma \gamma) \approx 2.34$\,KeV. This implies that the branching fraction $Br[\chi_{c0}(2P) \to J/\psi \, \omega]$ should be noticeably larger than the conservative lower bound $Br[\chi_{c0}(2P) \to J/\psi \, \omega] > 2.5\%$ which translates to the lower bound of 0.5\,MeV for the rate in Eq.(\ref{gam2}). Also, given that the measured~\cite{babar} product of the branching fractions in the $B$ meson decays,
$Br[B \to \chi_{c0}(2P) \, K] \, Br[\chi_{c0}(2P) \to J/\psi \, \omega] \approx (2 \,-  \, 3) \times 10^{-5}$ (or even as large as~\cite{belle} $7 \times 10^{-5}$), assuming a value of $Br[\chi_{c0}(2P) \to J/\psi \, \omega]$ near the indicated conservative lower limit would imply $Br[B \to \chi_{c0}(2P) \, K] \sim 10^{-3}$ which would thus be several times larger than similar branching fractions for the 1P charmonium states. Although not entirely excluded by the data, such behavior would be considerably challenging to explain. 

We thus conclude that a value of the coupling $g_\omega^2$ of order $10^{-2}$, though not understood theoretically, does not appear to be unnatural for charmonium. 

It may also be instructive to compare the relative strength of the coupling in the $\omega$ transitions in the charmonium and bottomonium sectors. The observed~\cite{belle2} rate of the transition $\Upsilon(5S) \to \chi_{b1} \, \omega$ corresponds to $g^2_\omega(5S \to 1P) \approx 3 \times 10^{-4}$, i.e. an order of magnitude smaller in the amplitude than the discussed coupling in charmonium. It is not clear however to what extent this transition is representative for bottomonium, since for the transitions from $\Upsilon(5S)$ the data indicate a strong violation of the heavy quark symmetry prediction (5 : 3)  for the relative yield of $\chi_{b2}$ and $\chi_{b1}$. Whatever the mechanism responsible for the enhanced spin symmetry breaking is (e.g. a contribution of meson-antimeson states~\cite{mv2}), it would also affect the strength of the coupling for the $\omega$ emission.
Perhaps, a more sensible comparison could be done using the data on the decays $\chi_{b1,b2}(2P) \to \Upsilon(1S) \, \omega$ for which the data on their relative rate do not contradict the heavy quark symmetry.  
However the absolute rate of the $\omega$ transitions from the $\chi_{bJ}(2P)$ bottomonium to $\Upsilon(1S)$ is unknown since the total widths of the $\chi_{bJ}(2P)$ states is not measured. Using the theoretical estimates~\cite{kr,gr} of the radiative and the total widths of the bottomonium $2P$ states one may deduce e.g. $\Gamma[\chi_{b1}(2P) \to \Upsilon(1S) \, \omega] \approx 1.5$\,keV, and thus estimate $g^2_\omega(2P \to 1S) \approx 2 \times 10^{-5}$. Clearly, these estimates point to a much weaker coupling for the $\omega$ transitions in bottomonium as compared to charmonium. Qualitatively, a weaker coupling for the $b \bar b$ system would be expected on general grounds, however at present we are not aware of any existing quantitative analysis, neither we can offer one ourselves.  

In summary, we find that a minimalistic description of the recent BESIII data on $e^+e^- \to \chi_{cJ} \, \omega$  in terms of the effect of the known resonance $\psi(4160)$ is quite compatible with the measurements of the cross section. This description implies that the coupling of the $\psi(4160)$ to the channels $\chi_{cJ} \, \omega$ is of the same order as in another known charmonium process $\chi_{c0}(2P) \to J/\psi \, \omega$. In both these cases the interaction of the $\omega$ meson is significantly stronger than in similar processes in bottomonium --- a behavior that is generally expected, but not yet described quantitatively.

The work of M.~B.~V.  is supported, in part, by the DOE grant DE-SC0011842.


\begin{thebibliography}{99}
\bibitem{bes} 
  M.~Ablikim {\it et al.}  [BESIII Collaboration],
  arXiv:1410.6538 [hep-ex].

\bibitem{pdg} 
  K.~A.~Olive {\it et al.}  [Particle Data Group Collaboration],
  Chin.\ Phys.\ C {\bf 38}, 090001 (2014).	
\bibitem{bes2} 
  M.~Ablikim {\it et al.}  [BES Collaboration],
  eConf C {\bf 070805}, 02 (2007)
  [Phys.\ Lett.\ B {\bf 660}, 315 (2008)]
  [arXiv:0705.4500 [hep-ex]].
	
\bibitem{belle} 
  K.~Abe {\it et al.}  [Belle Collaboration],
  Phys.\ Rev.\ Lett.\  {\bf 94}, 182002 (2005)
  [hep-ex/0408126].
	
\bibitem{cleo} 
  H.~Severini {\it et al.}  [CLEO Collaboration],
  Phys.\ Rev.\ Lett.\  {\bf 92}, 222002 (2004)
  [hep-ex/0307034].
	
\bibitem{belle2} 
  X.~H.~He {\it et al.}  [Belle Collaboration],
  Phys.\ Rev.\ Lett.\  {\bf 113}, 142001 (2014)
  [arXiv:1408.0504 [hep-ex]].
	
\bibitem{gottfried} 
  K.~Gottfried,
  Phys.\ Rev.\ Lett.\  {\bf 40}, 598 (1978).
	
\bibitem{mv0} 
  M.~B.~Voloshin,
  Nucl.\ Phys.\ B {\bf 154}, 365 (1979).
	
\bibitem{mv} 
  M.~B.~Voloshin,
  Mod.\ Phys.\ Lett.\ A {\bf 18}, 1067 (2003)
  [hep-ph/0304165].
	
\bibitem{mv2} 
  M.~B.~Voloshin,
  Phys.\ Rev.\ D {\bf 85}, 034024 (2012)
  [arXiv:1201.1222 [hep-ph]].
	
\bibitem{mllz} 
  L.~Ma, X.~H.~Liu, X.~Liu and S.~L.~Zhu,
  arXiv:1406.6879 [hep-ph].
	
\bibitem{babar} 
  P.~del Amo Sanchez {\it et al.}  [BaBar Collaboration],
  Phys.\ Rev.\ D {\bf 82}, 011101 (2010)
  [arXiv:1005.5190 [hep-ex]].
	
\bibitem{kr} 
  W.~Kwong and J.~L.~Rosner,
  Phys.\ Rev.\ D {\bf 38}, 279 (1988).
	
\bibitem{gr} 
  S.~Godfrey and J.~L.~Rosner,
  Phys.\ Rev.\ D {\bf 66}, 014012 (2002)
  [hep-ph/0205255].
\end{thebibliography}
\end{document}